\documentstyle[11pt,epsfig,amssymb,color, here]{article}

\begin{document}
\vspace*{-1.8cm}
\begin{flushright}
\flushright{\bf LAL 01-30}\\
\vspace*{-0.5cm}
%\hspace*{-12.5cm}
\flushright{July 2001}
\end{flushright}
\vskip 2.5 cm

\begin{center}
{\large\bf SEARCHES FOR LEPTOQUARKS, SQUARKS IN $\not\!\! R_p$ 
SUSY}\\
\vspace{0.2 cm}
{\large\bf AND EXCITED FERMIONS AT HERA}
\end{center}
\vskip 1.5 cm

\begin{center}
{\large\bf Zhiqing ZHANG}
\end{center}
\begin{center}
{\large\bf Laboratoire de l'Acc\'el\'erateur Lin\'eaire}\\
{IN2P3-CNRS et Universit\'e de Paris-Sud, BP 34, F-91898 Orsay Cedex}\\
{\it E-mail: zhangzq@lal.in2p3.fr\\
(On behalf of the H1 and ZEUS Collaborations)}
\end{center}

%\vspace*{0.5cm}

%\begin{abstract}

%%%%%%

\font\eightrm=cmr8

\bibliographystyle{unsrt} %for BibTeX - sorted numerical labels by
                          %order of first citation.

\arraycolsep1.5pt

% A useful Journal macro
\def\Journal#1#2#3#4{{#1} {\bf #2}, #3 (#4)}

% Some useful journal names
\def\NCA{\em Nuovo Cimento}
\def\NIM{\em Nucl.\ Instrum.\ Methods}
\def\NIMA{{\em Nucl.\ Instrum.\ Methods} A}
\def\NPB{{\em Nucl.\ Phys.} B}
\def\PLB{{\em Phys.\ Lett.} B}
\def\PRL{\em Phys.\ Rev.\ Lett.}
\def\PRD{{\em Phys.\ Rev.} D}
\def\ZPC{{\em Z.\ Phys.} C}
\def\PR{\em Phys.\ Rep.}
\def\EPJ{{\em Eur.\ Phys.\ J.} C}

% Some other macros used in the sample text
\def\st{\scriptstyle}
\def\sst{\scriptscriptstyle}
\def\mco{\multicolumn}
\def\epp{\epsilon^{\prime}}
\def\vep{\varepsilon}
\def\ra{\rightarrow}
\def\ppg{\pi^+\pi^-\gamma}
\def\vp{{\bf p}}
\def\ko{K^0}
\def\kb{\bar{K^0}}
\def\al{\alpha}
\def\ab{\bar{\alpha}}
\def\be{\begin{equation}}
\def\ee{\end{equation}}
\def\bea{\begin{eqnarray}}
\def\eea{\end{eqnarray}}
\def\CPbar{\hbox{{\rm CP}\hskip-1.80em{/}}}%temp replacemt due to no font

%%%%%%%%%%%%%%%%%%%%%%%%%%%%%%%%%%%%%%%%%%%%%%%%%%%%%%%%%%%%%%%%%%%%%%%%
%%BEGINNING OF TEXT                           
%%%%%%%%%%%%%%%%%%%%%%%%%%%%%%%%%%%%%%%%%%%%%%%%%%%%%%%%%%%%%%%%%%%%%%%%

%\begin{document}
%\vspace{-5cm}
%{\flushright \large {\bf LAL 01-30}\\}
%\vspace{-3mm}
%{\flushright July 2001\\}
%\title{}

%\author{Z. ZHANG}

%\address{LAL, Univ.\ Paris-Sud et IN2P3/CNRS,\\ BP 34, 91898 Orsay Cedex, 
%FRANCE\\ E-mail: zhangzq@lal.in2p3.fr\\
%(On behalf of the H1 and ZEUS Collaborations)}


%\maketitle
%\abstracts{
\begin{abstract}
Various searches for leptoquarks, scalar quarks in 
$R_p$-violating supersymmetric models, and excited fermions 
performed by the HERA experiments H1 and ZEUS are reviewed. No evidence 
for new particle production was observed from data collected by both 
experiments since 1994 in either electron-proton and positron-proton 
collisions. Stringent limits derived on the masses and couplings of 
these new particles
%, as well as future prospects 
are compared whenever appropriate with those
from LEP, the Tevatron and low energy experiments.
\end{abstract}

\section{Introduction}
%\subsection{Producing the Hard Copy}\label{subsec:prod}
The $ep$ collider HERA, which provides both baryonic ($B$) and leptonic ($L$) 
quantum numbers in the initial state, is ideally suited to search for 
new particles possessing couplings to a lepton-quark pair. 
Such particles could be leptoquarks (LQs) or scalar quarks 
(squarks, $\tilde{q}$). LQs are color-triplet bosons which appear
in many extensions of the Standard Model (SM).
%such as Grand Unified Theories~\cite{gut} and Superstring inspired $E_6$ models~\cite{e6}, 
%and in some Compositeness~\cite{composite} and Technicolor~\cite{tc} models. 
Squarks, the scalar supersymmetric (SUSY) partners of quarks, also couple to
a lepton-quark pair in models which violate $R$-parity. The $R$-parity, 
defined as $R_p=(-1)^{F+2S}$ with $F (=3B+L)$ and $S$ being respectively 
the fermion number and the spin, is a discrete symmetry. 
HERA also provides ideal conditions to look for
excited fermions ($e^\ast$, $\nu^\ast$, and $q^\ast$) of the first generation.
The existence of these excited states would provide clear evidence for 
fermion substructure.

This note briefly reviews all recent searches at HERA based on three
independent data samples taken since 1994.
The most recently published results were obtained using data collected
from 1994 to 1997 at a center-of-mass energy
$\sqrt{s}(=\!\sqrt{4E_eE_p})$ of 300\,GeV by colliding {\it positrons} of
energy 27.5\,GeV with protons of 820\,GeV.
The results from 1998-1999 {\it electron}-proton collisions at a slightly
higher energy $\sqrt{s}\simeq 320$\,GeV (due to an increase of
the proton beam energy to 920\,GeV) are mostly still in their preliminary form.
The new $e^+p$ data taken in 1999-2000 at $\sqrt{s}=320$\,GeV are being 
analyzed and first results are presented here.
The searches for LQs and squarks in $R_p$-violating ($\not\!\! R_p$) SUSY
are presented, respectively, in Secs.\,\ref{sec:lq} and \ref{sec:susy}.
Sec.\,\ref{sec:exfermion} describes the searches for excited
fermions, followed by Sec.\,\ref{sec:summary} with a summary and an outlook.

\section{Searches for Leptoquarks}\label{sec:lq}
At HERA, LQs could be resonantly produced in the $s$-channel by the fusion
of the initial state lepton with a quark from the proton or virtually
exchanged in the $u$-channel. LQs can decay to $e+q$ and $\nu +q$
such that the amplitudes of the $s$- and $u$-channel diagrams 
interfere with those from deep inelastic scattering (DIS) processes
%with either a photon ($\gamma$) or a $Z^0$ boson and
%a $W$ boson exchanged in the $t$-channel respectively in 
of the neutral current
(NC) and charged current (CC) interactions.

In general, the $s$-channel contribution dominates for LQ masses up to
$\sqrt{s}$. The production cross-section can be written to,
a good approximation, as
%\begin{equation}
$\sigma_{\rm LQ}=(J+1)\frac{\pi\lambda^2}{4s}q\!\left(x=\frac{M^2_{\rm LQ}}{s}\right)$
%\end{equation}
where $J=0$ and 1 respectively for scalar and vector LQs, $\lambda$ is 
the Yukawa coupling at the LQ-$e$-$q$ vertex, and $q(x)$ is the parton
density. The $u$-channel and interference contributions cannot produce
a resonance peak and are only important for $M_{\rm LQ}\gtrsim \sqrt{s}$.

The phenomenological model proposed by Buchm\"uller-R\"uckl-Wyler 
(BRW)~\cite{brw} describes 14 LQs, of which 7 have $F=0$ and 7 have $F=2$. 
The $e^+p$ collisions provide the best sensitivity to the former 
since the fusion involves a quark instead of an antiquark.
This is in contrast to the $e^-p$ collisions where the $F=2$ states 
are best probed.

The final states of LQ decay are identical to NC and CC DIS processes.
On the other hand,
since the angular distributions of the decay products of a scalar or
vector resonance are different from those of DIS (in particular for NC),
a mass dependent angular cut, or equivalently a cut in $y$,
is applied to maximize the signal significance. An excess
of events was originally observed in the mass and $Q^2$
distributions of the NC channel in the 1994-1996 $e^+p$ data 
corresponding to an integrated luminosity of about 15\,pb$^{-1}$ per
experiment. However such an excess was not confirmed by later higher
statistics samples of $e^+p$ data ($\sim 20$\,pb$^{-1}$ in 1997 and 
$\sim 70$\,pb$^{-1}$ in 1999-2000) and by the $e^-p$ data 
($\sim 15$\,pb$^{-1}$).

\begin{figure}[h]
\begin{center}
\begin{picture}(50,140)
%\rule{5cm}{0.2mm}\hfill\rule{5cm}{0.2mm}
%\vskip 2.5cm
%\rule{5cm}{0.2mm}\hfill\rule{5cm}{0.2mm}
%\put(-120,-50){\epsfig{figure=lq_f0.eps,width=10cm,height=7cm}}
\put(-120,-50){\epsfig{figure=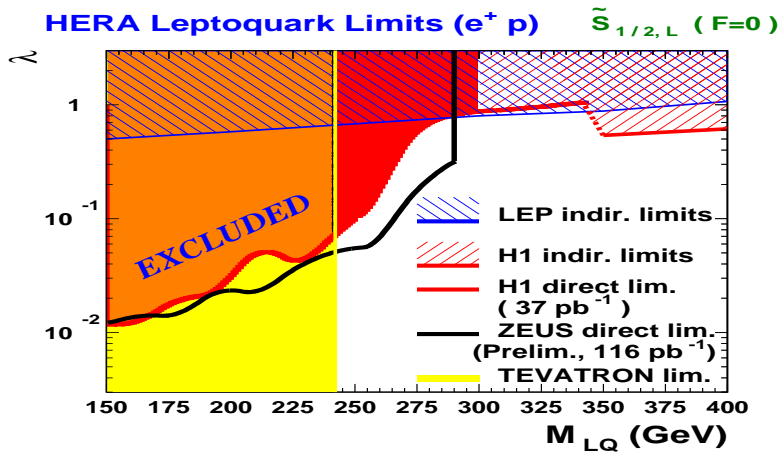,width=11cm,height=8cm}}
\end{picture}
\end{center}
\vspace{-2mm}
\caption{Upper exclusion limits at 95\% CL on the Yukawa coupling $\lambda$
as a function of the LQ mass for a left-handed scalar $\tilde{S}_{1/2,L}$ with
isospin 1/2 and fermion number $F=0$.\label{fig:lqe+}}
\end{figure}

%Assuming Poisson distributions for the SM background expectations and
%for the signal, an upper limit on the number of events coming from LQ
%production is obtained using a standard Bayesian prescription. The limit
%on the number of signal events is then translated into an upper bound
%on the LQ cross-section, which in turn leads to constraints e.g.\ on the
%coupling $\lambda$.
Since no evidence for LQ production was observed, upper limits were
derived on the LQ cross-section and coupling $\lambda$.
As an example, upper limits at 95\% confidence level (CL) 
obtained for a LQ state $\tilde{S}_{1/2,L}$ are shown
in Fig.~\ref{fig:lqe+}.
\vskip 0.5cm
%%%%fig 1

The H1 direct limit~\cite{h1lq9497} for masses below 300\,GeV was derived 
from the 1994-1997 $e^+p$ data. 
By taking properly into account the $u$-channel and interference
contributions, a sensitivity to coupling values $\lesssim 1$ was established 
for masses up to 400\,GeV. The better sensitivity at higher masses was achieved
by a contact interaction analysis~\cite{h1ci} where part of the 1999-2000 $e^+p$ data
was combined with those of 1994-1997. The preliminary ZEUS limit, presented
for the first time in this workshop, was obtained
using the full $e^+p$ data. The higher sensitivity at 
$M_{\rm LQ}\gtrsim 250$\,GeV is largely due to the increased $\sqrt{s}$. 
In comparison with limits from LEP~\cite{leplq} and 
the Tevatron~\cite{tevatronlq}, HERA thus provides
the best sensitivity at the intermediate and high mass range.
%ce\begin{figure}[tb]
%ce\begin{center}
%ce\begin{picture}(50,140)

%\put(-120,-50){\epsfig{figure=lq_f0.eps,width=11cm,height=8cm}}
%\end{picture}
%\end{center}
%\caption{Upper exclusion limits at 95\% CL on the Yukawa coupling $\lambda$
%as a function of the LQ mass for a left-handed scalar $\tilde{S}_{1/2,L}$ with
%isospin 1/2 and fermion number $F=0$.\label{fig:lqe+}}
%\end{figure}

\begin{figure}[h]
\begin{center}
\begin{picture}(50,145)
\put(-120,-45){\epsfig{figure=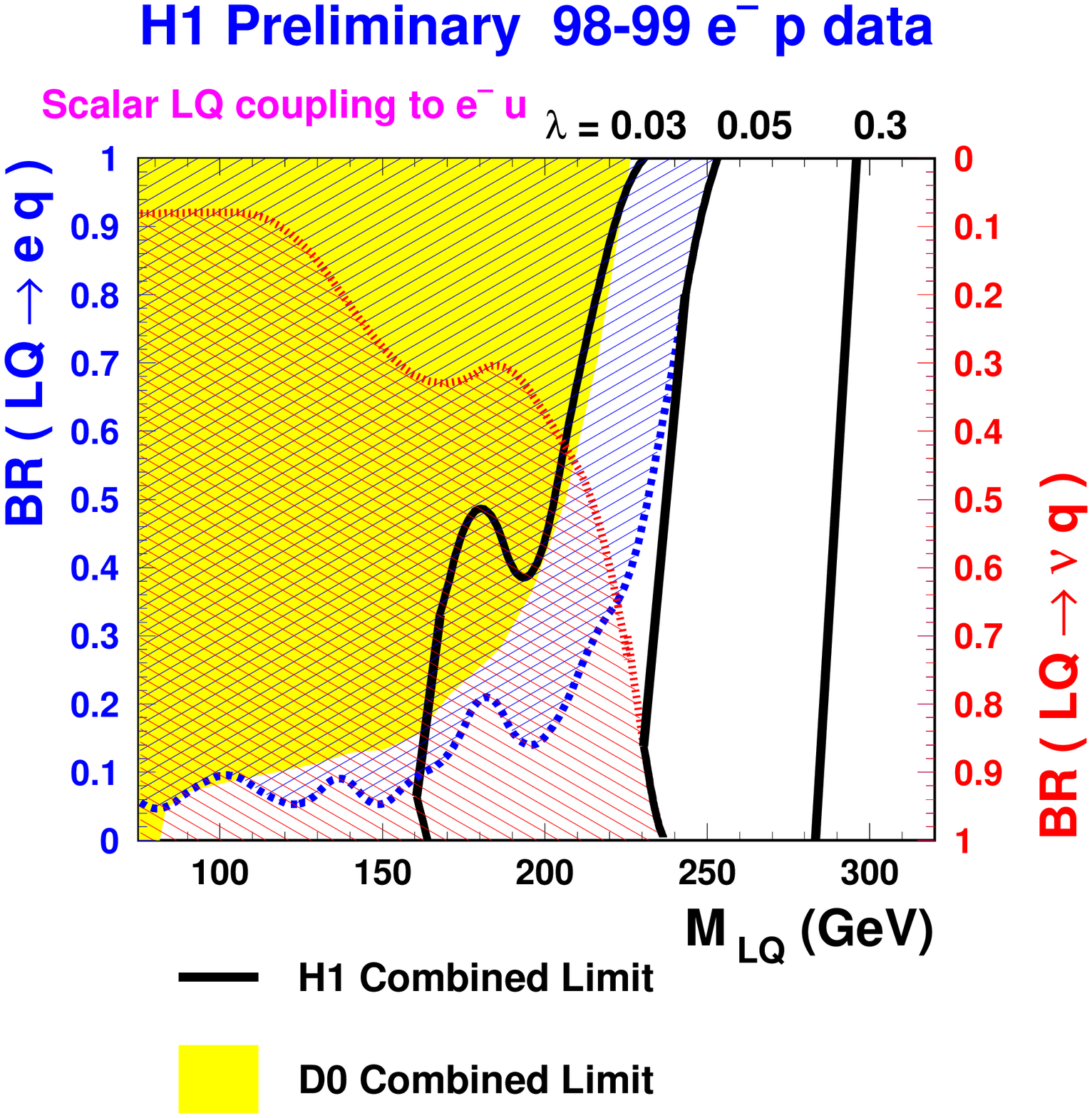,width=10cm,height=7cm}}
\put(-70,-34){\fcolorbox{white}{white}{\textcolor{white}{\rule{43mm}{8mm}}}}
\end{picture}
\end{center}
\vspace{2mm}
\caption{Mass dependent exclusion limits at 95\% CL in generic LQ models
for a scalar LQ which couples to $e^- u$ shown as the dashed (dotted) curve 
as a function of the branching ratio
BR$({\rm LQ}\rightarrow eq)$ (BR$({\rm LQ}\rightarrow \nu q)$) for 
an example coupling $\lambda=0.05$. The combined limits are represented 
with solid curves for three coupling values. The region excluded by D0 is 
indicated with the shaded area.
\label{fig:lqe-}}
\end{figure}

In generic LQ models, the branching ratios of LQ decays in the NC and CC
DIS -like
modes are free parameters in contrast to the BRW model where they are fixed
to 1, 1/2, or 0. If the LQ decays into $e+q$ and $\nu +q$ only, the combined
preliminary H1 results obtained from the 1998-1999 $e^-p$ data are shown in
Fig.~\ref{fig:lqe-}. A similar analysis has been performed by ZEUS 
using the 1994-1997 $e^+p$ data~\cite{zeuslq9497}.
The combined bounds are largely independent of the individual
branching ratios. As soon as $\lambda$ exceeds $\sim 0.03$, 
these limits extend considerably beyond the region excluded by D0, 
as represented by the shaded domain. 

\section{Searches for Squarks in $\not\!\! R_p$ SUSY}\label{sec:susy}
SUSY is one of the most likely ingredients for a theory beyond the SM.
In particular, the Minimal Supersymmetric extension of the SM (MSSM) 
describes all experimental data just as well as the SM. 
%, and in addition
%it offers among its appealing consequences solutions for the
%cancellation of quadratic divergences occurring in the scalar Higgs
%sector of the SM and models beyond the SM.
%
The most general SUSY theory which preserves gauge invariance of the 
SM allows for $\not\!\! R_p$ Yukawa couplings $\lambda$, 
$\lambda^{\prime}$, $\lambda^{\prime\prime}$ between
one squark or slepton and two SM fermions:
$W_{\not\! R_p}=\lambda L_iL_j\overline{E}_k+
\lambda^\prime L_iQ_j\overline{D}_k+
\lambda^{\prime\prime} \overline{U}_i\overline{D}_j\overline{D}_k$,
where $i,j,k=1,2,3$ are generation indices, $L_i(Q_i)$ are
the left-handed lepton (quark)
%$SU(2)_L$
doublet superfields and
$\overline{E}_i(\overline{D}_i,\overline{U}_i)$ are the right-handed
electron (down and up quark)
%$SU(2)_L$
singlet superfields. Of particular
interest for HERA are the $\not\!\! R_p$ terms
$\lambda^\prime L_iQ_j\overline{D}_k$. The squarks at HERA are
thus singly produced in the $s$-channel with masses up to the
kinematic limit.

In the case where both production and decay occur through the same
$\lambda^\prime_{1jk}$, the squarks in $\not\!\! R_p$-SUSY behave as scalar
LQs and the constraints obtained on LQs are also applicable for squarks.
%There is however an interesting difference,
%namely the exclusion phase space covered by HERA data can be relatively
%larger for
%squarks than for LQs. This is because the mass constraints~\cite{tevatronlq}
%of 242\,GeV (205\,GeV)
%from the Tevatron correspond to a branching ratio of the new particle into $e+q$ of 1 (0.5),
%which can be naturally small in $\not\!\! R_p$-SUSY framework given  the competition
%with gauge decay modes of the squarks.
There are however other decay modes in which a squark decays to a quark
and a gaugino (chargino or neutralino) with $R_p$-conserving gauge couplings.
Because of the gauge decay modes,
the resulting constraints on the mass and coupling of squarks depend
on the various parameters in the SUSY phase space. 
The search sensitivity is improved considerably when the gauge decay
modes are combined with those of $\not\!\!\! R_p$ decays.
\newpage

With $e^+p$ collisions, HERA is most sensitive to the couplings 
$\lambda^\prime_{1j1}$ amongst the nine possible couplings 
$\lambda^\prime_{1jk}$, where mainly $\tilde{u}^j_L$
squarks are produced via processes involving a valence $d$ quark\footnote{On the contrary,
the $e^-p$ data are well suited to probe couplings $\lambda^\prime_{11k}$ and
$\tilde{d}^k_R$ squarks.}. Mass dependent limits~\cite{herasusy}
on $\lambda^\prime$ were derived
by both H1 and ZEUS within the MSSM model and by H1 within
%in the framework of
the minimal supergravity model, a more constrained SUSY model. 
The model dependence of the results was studied in detail by performing 
a scan of the MSSM parameters and was found to be small. The HERA direct 
searches improve the best indirect limits on $\lambda_{1j1}, j=2,3$ from
atomic parity violation measurements by a factor of up to 3. 
In a large part of the MSSM parameter space covered 
by the scan, the existence of squarks coupling to a $e^+d$ pair with masses 
up to 260\,GeV is excluded at 95\% CL for a coupling of 
electromagnetic strength.
\vspace{2mm}
\section{Searches for Excited Fermions}\label{sec:exfermion}
At HERA, excited fermions ($f^\ast$) could be singly produced 
via the $t$-channel exchange
of a gauge boson, and would subsequently decay into a SM fermion and a boson.
Collider searches are generally interpreted in the framework of the
phenomenological model~\cite{hagiwara}, where the interactions of $f^\ast$ 
with a SM fermion and an electroweak boson (a gluon), 
$L_{\rm eff}=\frac{1}{2\Lambda}F^\ast_R\sigma^{\mu\nu}
\left[gf\frac{\tau^a}{2}W^a_{\mu\nu}+\right.$$g^\prime f^\prime\frac{Y}{2}B_{\mu\nu}+$
$\left.g_sf_s\frac{\lambda^a}{2}G^a_{\mu\nu}\right]F_L+h.c.$,
are parameterized via the compositeness scale $\Lambda$ and
relative couplings $f$ and $f^\prime$ ($f_s$).

The search has investigated the decays of $f^\ast$ into $\gamma$,
$Z$, and $W$, followed by the subsequent decay of the boson into
$e$, $\mu$, $\nu$, or hadrons. No deviation from the SM predictions
has been observed, which leads to constraints on the considered
model. Based on the 1994-1997 $e^+p$ data at $\sqrt{s}=300$\,GeV,
upper limits\footnote{The values given here are from H1~\cite{h1ef}, similar
limits have also been obtained by ZEUS~\cite{zeusef}.} 
for $f/\Lambda$ (with $f=f^\prime$)\footnote{The case 
$f=-f^\prime$ was not considered since the production cross-section
of the $e^\ast$ would be very small due to the vanishing coupling constant.}
ranging from $7\times 10^{-4}$ to $10^{-2}\,{\rm GeV}^{-1}$ were
obtained for an $e^\ast$ mass ranging from 50\,GeV to 250\,GeV.
For an $\nu^\ast$ mass ranging from 50 to 200\,GeV, the values of
the limits for $f/\Lambda$ (with $f=-f^\prime$) vary between 
$3\times 10^{-3}$ and $10^{-1}\,{\rm GeV}^{-1}$. The limits
when $f=f^\prime$ were less stringent due to the absence of the dominant decay 
$\nu^\ast\rightarrow \nu \gamma$. Assuming $f/\Lambda=1/M_{l^\ast}$, masses
below 223 and 114\,GeV are excluded at 95\% CL, respectively, 
for the $e^\ast (f=f^\prime)$ and $\nu^\ast(f=-f^\prime)$ production.
These limits extend beyond those obtained at LEP where the constraints
are limited either to beam energies for $l^\ast$ produced in pairs or
to the center-of-mass energies for $l^\ast$ produced singly.
Limits for the $q^\ast$ on $f/\Lambda$ assuming $f=f^\prime$ and $f_s=0$ (i.e.
only electroweak couplings) vary between $9\times 10^{-4}$
and $2\times 10^{-2}\,{\rm GeV}^{-1}$ for $q^\ast$ masses ranging 
from 50 to 250\,GeV.
The HERA limits are complementary to those from the Tevatron where
stringent bounds were set on $q^\ast$ produced in $qg$ fusion via
the coupling $f_s$.

Using the 1998-1999 $e^-p$ data, preliminary limits~\cite{h1ef,zeusef}
on $f/\Lambda$ for $\nu^\ast$ 
have been obtained by both HERA experiments, which improve
significantly the limit based on the $e^+p$ data, e.g.\ by H1 
from 114\,GeV to 150\,GeV for $f=-f^\prime$, due to the much 
higher production cross-section.
\vspace{2mm}
\section{Summary and Outlook}\label{sec:summary}
New particle production of leptoquarks, scalar quarks, and excited fermions
has been extensively searched for at HERA. Many final results based on
the 1994-1997 $e^+p$ data were published. Some preliminary results
using the 1998-1999 $e^-p$ are available, which either complement
the searches for new particle types (e.g.\ $F=2$ versus $F=0$ LQs 
probed respectively with $e^-p$ and $e^+p$ data) or extend substantially 
the limits (e.g.\ for $\nu^\ast$).\linebreak
\noindent
 Higher sensitivity has been obtained using
the 1999-2000 $e^+p$ data on the Yukawa coupling of the LQ and such
an improvement is also expected for searches for squarks and excited
fermions. HERA limits were found to be competitive and complementary 
in comparison with those from other high energy machines and 
low energy experiments.

The HERA machine and both the H1 and ZEUS detectors are being upgraded. 
A factor of about 5 increase in the luminosity and the improved detectors 
after the upgrade will provide new particle hunters with new
and exciting opportunities in the next years before the era of the LHC.

%\section*{Acknowledgments}
%This is where one places acknowledgments for funding bodies etc.  
%Note that there are no section numbers for the Acknowledgments, 
%Appendix or References.

%\section*{References}

\end{document}